\documentclass[aps,prl,twocolumn,superscriptaddress,showpacs]{revtex4}

\usepackage{graphicx}

\begin{document}

\title{Coloring random graphs}

\author{R. Mulet}
\affiliation{International Centre for Theoretical Physics, Strada
Costiera 11, P.O. Box 586, 34100 Trieste, Italy}
\affiliation{Henri-Poincar\'e-Chair of Complex Systems and
Superconductivity Laboratory, Physics Faculty-IMRE, University of 
Havana, La Habana, CP 10400, Cuba}

\author{A. Pagnani}
\affiliation{International Centre for Theoretical Physics, Strada
Costiera 11, P.O. Box 586, 34100 Trieste, Italy}
\affiliation{Dipartimento di Fisica, INFM and SMC, Universit\`a di 
Roma ``La Sapienza'', P.le Aldo Moro 2, 00185 Roma, Italy}

\author{M. Weigt}
\affiliation{International Centre for Theoretical Physics, Strada
Costiera 11, P.O. Box 586, 34100 Trieste, Italy}
\affiliation{Institute for Theoretical Physics, University of
G\"ottingen, Bunsenstr. 9, 37073 G\"ottingen, Germany}

\author{R. Zecchina}
\affiliation{International Centre for Theoretical Physics, Strada
Costiera 11, P.O. Box 586, 34100 Trieste, Italy}

\date{\today}

\begin{abstract}

We study the graph coloring problem over random graphs of finite
average connectivity $c$. Given a number $q$ of available colors, we
find that graphs with low connectivity admit almost always a proper
coloring whereas graphs with high connectivity are
uncolorable. Depending on $q$, we find the precise value of the
critical average connectivity $c_q$.  Moreover, we show that below
$c_q$ there exist a clustering phase $c\in [c_d,c_q]$ in which ground
states spontaneously divide into an exponential number of clusters and
where the proliferation of metastable states is responsible for the
onset of complexity in local search algorithms.

\end{abstract}

\pacs{89.20.Ff, 75.10.Nr, 05.70.Fh, 02.70.-c}

\maketitle

The Graph Coloring problem (COL) is a very basic and famous problems
in combinatorics \cite{GaJo} and in statistical physics \cite{WU}.
Given a graph, or a lattice, and given a number $q$ of available
colors, the problem consists in finding a coloring of vertices such
that no two neighboring vertices have the same color. The minimally
needed number of colors is the {\it chromatic number} of the graph.

For planar graphs there exists a famous theorem~\cite{four_colors}
showing that four colors are sufficient, and that a coloring can be
found by an efficient algorithm. On the contrary, for general graphs
the problem is computationally hard to solve: already in 1972 it was
shown that Graph Coloring is NP-complete \cite{Karp72} which means,
roughly speaking, that the time required for determining the existence
of a proper coloring grows exponentially with the graph size.

In modern computer science, graph coloring is taken as one of the most
widely used benchmarks for the evaluation of algorithm performance
\cite{satlib}. The interest in coloring stems from the fact that many
real-world combinatorial optimization problems have component
sub-problems which can be easily represented as coloring
problems. For instance, a classical application is the scheduling of
registers in the central processing unit of computers.  All variables
manipulated by the program are characterized by ranges of times during
which their values are left unchanged.  Any two variables that change
during the same time interval cannot be stored in the same register.
One may represent the overall computation by constructing a graph
where each variable is associated with a vertex and edges are placed
between any two vertices whose corresponding variables change during
the same time interval.  A proper coloring with a minimal number of
colors of this graph provides an optimal scheduling for registers: two
variables with the same color will not be connected by an edge and so
can be assigned to the same register (since they change in different
time intervals).

The $q$-coloring problem of random graphs represents a very active
field of research in discrete mathematics which constitutes the
natural evolution of the percolation theory initiated by Erd\"os and
R\`enyi in the 50's \cite{Erdos_Renyi}.  One point of contact between
computer science and random graph theory arises from the observation
that, for large random graphs, there exists a critical average
connectivity beyond which the graphs become uncolorable with
probability tending to one as the graph size goes to infinity. This
transition will be called the $q$-COL/UNCOL transition throughout this
paper. The precise value of the critical connectivity depends of
course on the number $q$ of allowed colors and on the ensemble of
random graphs under consideration.  Graphs generated close to their
critical connectivity are extraordinarily hard to color and therefore
the study of critical instances is at the same time a well posed
mathematical question as well as an algorithmic challenge for the
understanding of the onset of computational complexity \cite{AI,TCS}.
The notion of computational complexity refers to worst-case instances
and therefore results for a given ensemble of problems might not be of
direct relevance.  However, on the more practical side, algorithms
which are used to solve real-world problems display a huge variability
of running times and a theory for their typical-case behavior, on
classes of non-trivial random instances, constitutes the natural
complement to the worst-case analysis. Similarly to what happens for
other famous combinatorial problem, e.g. the satisfiability problem of
Boolean formulae, critical random instances of $q$-coloring (or
polynomial mappings to other NP-complete problems) are a popular
test-bed for the performance of search algorithms~\cite{satlib}.

In physics $q$-coloring has a direct interpretation as a
spin-glass model. A proper coloring of a graph is a zero-energy
ground state configuration of a Potts anti-ferromagnet with $q$-state
variables. For most lattices this system is frustrated and displays
many equilibrium and out-of-equilibrium features of glasses
(`Potts glass').

Here we focus on the $q$-coloring problem (or Potts anti-ferromagnet)
over random graphs of finite average connectivity, given by the ${\cal
G}_{N,p}$ ensemble (graphs composed of $N$ vertices with edge
probability $p$ for every pair of vertices).  In the relevant limit of
finite connectivity we have to take $p=c/N$ which leads to random
graphs with a Poissonian connectivity distribution of mean $c$.

Two types of questions can be asked.  One type is algorithmic, i.e.
finding an algorithm that decides whether a given graph is
colorable. The other type is more theoretical and amounts to asking
whether a typical problem instance is colorable or not and what is
the typical structure of the solution space.  Here we address the
latter question using the so called cavity method~\cite{MePa}.

Let us start with reviewing some known results on the $q$-COL/UNCOL
transition on random graphs. One of the first important
finite-connectivity results was obtained by Luczak about one decade
ago \cite{Lu}. He proved that the threshold asymptotically grows like
$c_q \sim 2q\ln q$ for large numbers of colors, a result, which up to
a pre-factor coincides with the outcome of a replica calculation on
highly connected graphs \cite{KaSo} ($p={\cal O}(1)$ for large $N$).
For fixed number $q$ of colors, all vertices with less than $q$
neighbors, i.e. of {\it degree} smaller $q$, can be colored for sure.
The hardest to color structure is thus given by the maximal subgraph
having minimal degree at least $q$, the so-called $q$-core. Pittel,
Spencer and Wormald \cite{PiSpWo} showed that the emergence of a
2-core coincides with random graph percolation at $c=1$
\cite{Erdos_Renyi}, and is continuous. For $q\geq 3$, however, the
$q$-core arises discontinuously: For $q=3$ they found, e.g., that the
core emerges at $c\simeq 3.35$ and immediately contains about 27\% of
all vertices. The existence of this core is, however, not sufficient
for uncolorability: The best lower bound for the 3-COL/UNCOL
transition is 4.03 \cite{AcMo1}, numerical results predict a threshold
of about 4.7 \cite{Numerical}. The currently best rigorous upper bound
is 4.99 \cite{AcMo2}. Most recently, a replica symmetric analysis of
the problem has been performed~\cite{MoSa}. The resulting threshold
5.1 exceeds, however, the rigorous bound, and one has to go beyond
replica symmetry. At the level of one-step replica-symmetry breaking
(1RSB) we are able to calculate a threshold value $c_3 \simeq 4.69$
which we believe to be exact \cite{note:rsb}. We also describe the
solution space structure which undergoes a clustering transition at
$c_d \simeq 4.42$.

As stated above, the question if a given graph is $q$-colorable
is equivalent to the question if there are zero-energy ground states
of the anti-ferromagnetic $q$-state Potts model defined on the same
graph. Denoting the set of all edges by $E$, the problem can thus
be described by the Hamiltonian
\begin{equation}
\label{eq:H}
H = \sum_{(i,j)\in E} \delta(\sigma_i,\sigma_j)
\end{equation}
with $\delta(\cdot,\cdot)$ denoting the Kronecker symbol. The spins
$\sigma_i, \ i=1,...,N,$ are allowed to take the $q$ values
$\{1,...,q\}$. This Hamiltonian counts the number of edges being
colored equally on both extremities, a proper coloring of the graph
thus has energy zero. In this letter we apply the cavity method in a
variant recently developed for finite-connectivity graphs directly at
zero temperature \cite{MePaZe,MePa2,MeZe}. This approach consists of 
a self-consistent iterative scheme which is believed to be exact over
locally tree-like graphs, like the ones we consider here. It includes
the possibility of dealing with the existence of many pure states. One
has to first evaluate the energy shift of the system due to the
addition of a new new spin $\sigma_0$. Let us assume for a moment that
the new spin is only connected to a single spin, say $\sigma_1$, in
the pre-existing graph. Before adding the new site 0, the ground-state
energy of the system with fixed $\sigma_1$ can be expressed as
\begin{equation}
\label{eq:Es1}
E_0(\sigma_1) = A - \sum_{\tau=1}^q h_\tau^1 \delta(\tau,\sigma_1)
\end{equation}
where we have introduced the effective field $\vec
h^1=(h_1^1,...,h_q^1)$.  Note that a $(q-1)$-dimensional field would
be sufficient since one of the $q$ fields above can be absorbed in
$A$. We, however, prefer to work with $q$ field components in order to
keep evident the global color symmetry. Connecting $\sigma_0$ to
$\sigma_1$, and calculating the minimal energy of the enlarged graph
with fixed $\sigma_0$, this reads
\begin{eqnarray}
\label{eq:Es0}
E(\sigma_0) &=&\min_{\sigma_1} \left( A - \sum_{\tau=1}^q h_\tau^1 
  \delta(\tau,\sigma_1) + \delta(\sigma_0,\sigma_1) \right)\nonumber\\
&=& A - \omega(\vec h ^1) - 
       \sum_{\tau=1}^q u_\tau(\vec h^1) \delta(\tau,\sigma_0)
\end{eqnarray}
with
\begin{eqnarray}
\label{eq:u}
\omega(\vec h) &=& -\min ( -h_1,..., -h_q )\\
u_\tau(\vec h) &=& 
\left\{
\begin{array}{rl}
-1 & \mbox{if} -h_\tau < -h_1,..,-h_{\tau-1},-h_{\tau+1},..,-h_q\\
0 & \mbox{else}
\end{array}
\right.\nonumber
\end{eqnarray}
The field $\vec u(\vec h^1)$ has at most one 
non-zero component, which takes the value $-1$, i.e. 
$\vec u(\vec h^1) \in \{ \vec 0, -\vec e_1,...,- \vec e_q \}$ with 
$\vec e_\tau$ denoting a unit vector in direction $\tau$. 

If the new spin $\sigma_0$ is connected to $d$ sites with fields $\vec
h^1,...,\vec h^d$, and if these spins were previously uncorrelated
(which is the case inside one pure state, cf. the discussion in
\cite{MePa}), the propagated fields can be linearly superposed, $\vec
h^0 = \sum_{i=1}^d \vec u (\vec h ^i)\ $.
Note that the fields never become positive, which reflects the
anti-ferromagnetic character of the model. Colors are suppressed by
neighbors carrying this color, they can be favored only by suppressing
all other colors.  If there would be a single state (replica
symmetry), every link $(i,j)$ would carry two propagated fields $\vec
u_{i\to j}$ and $\vec u_{j\to i}$ which are determined
self-consistently. In case of multiple states, these fields fluctuate
from state to state and have to be characterized by a full
distribution $Q_{i\to j}(\vec u)$, cf. \cite{MePa,MeZe}. Due to the 
global color symmetry, each of these takes the form
\begin{equation}
\label{eq:Q}
Q_{i\to j}(\vec u) = (1-q\eta_{i\to j}) \delta(\vec u) + 
\eta_{i\to j} \sum_{\tau=1}^q \delta(\vec u + \vec e_\tau)
\end{equation}
and can thus be fully described by the probability $\eta_{i\to j}$
that any of the colors of vertex $j$ is forbidden by edge $(i,j)$.
Denoting the histogram of all $\eta_{i\to j}$ by $\rho(\eta)$, the
self-consistency equation for the distribution of the $Q_{i\to j}(\vec
u)$ can be reduced to a simple equation for $\rho(\eta)$ \cite{note},
\begin{eqnarray}
\label{eq:rho}
\rho(\eta) &=& e^{-c} \sum_{d=0}^\infty \frac{c^d}{d!} \int_0^{1/q}
d\eta_1\cdots d\eta_d\ \rho(\eta_1)\cdots \rho(\eta_d) \nonumber\\
&&\times \delta( \eta - f_d(\eta_1,...,\eta_d) )
\end{eqnarray}
where $f_d$ is simply given by
\begin{equation}
\label{eq:newrho}
f_d(\eta_1,...,\eta_d) = 
\frac{\sum_{l=0}^{q-1} (-1)^l {q-1 \choose l} \prod_{i=1}^d
[1-(l+1)\eta_i] }
{\sum_{l=0}^{q-1} (-1)^l {q \choose l+1} \prod_{i=1}^d
[1-(l+1)\eta_i] }
\end{equation}
This equation resembles a replica-symmetric self-consistent equation
and can be solved numerically using a population dynamical algorithm:
We start with an initial population $\eta_1,...,\eta_{\cal N}$ of size
${\cal N}$ which can be easily chosen to be as large as $10^6$ to
generate high-precision data.  This population is updated by iterating
the following steps until convergence: (I) Randomly draw a number $d$
from the Poisson distribution $e^{-c} c^d/d!$; (II) Randomly select
$d+1$ indices $i_0,i_1,...,i_d$ from $\{1,...,{\cal N}\}$; (III)
Update the population by replacing $\eta_{i_0}$ by
$f_d(\eta_{i_1},...,\eta_{i_d})$.

One obvious solution of Eq. (\ref{eq:newrho}) is the paramagnetic
solution $\delta(\eta)$. For small average connectivities $c$ it is
even the only one. The appearance of a non-trivial solution coincides
with a clustering transition of ground states into an exponentially
large number of extensively separated clusters. In spin-glass theory,
this transition is called dynamical.  Still, $\rho(\eta)$ will contain
a non-trivial peak in $\eta=0$ due to small disconnected subgraphs,
dangling ends etc. The weight $t$ of this peak can be computed
self-consistently from
\begin{equation}
\label{eq:qcore}
t= e^{-(1-t)c} \sum_{l=0}^{q-2} \frac{(1-t)^lc^l}{l!}\ .
\end{equation}
This equation is quite interesting, since a non-trivial solution forms
a necessary condition for Eq. (\ref{eq:rho}) to have a non-trivial
solution. In fact \cite{PiSpWo}, the fraction of edges in the $q$-core
is given by $(1-t_{min})$ with $t_{min}$ being the smallest positive
solution of Eq. (\ref{eq:qcore}). Thus, we also find that the
existence of an extensive $q$-core is necessary for a non-trivial
$\rho(\eta)$, and forms a lower bound for the $q$-COL/UNCOL
transition.

\begin{figure}[htb]
\includegraphics[width=0.94 \columnwidth]{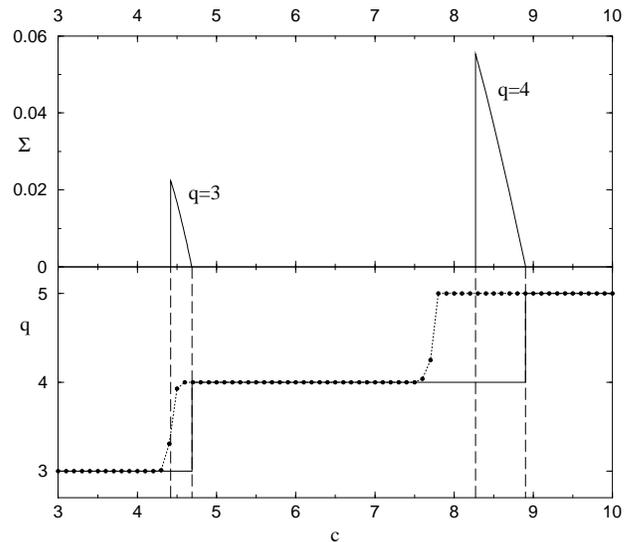}
\caption{Top: Complexity $\Sigma(c)$ vs. average connectivity for
$q=3$ and $q=4$. Non-zero complexity appears discontinuously at
the dynamical threshold $c_d$, and goes down continuously to
zero at the $q$-COL/UNCOL transition. The curves are calculated 
using the population-dynamical solution for $\rho(\eta)$ with
population size ${\cal N}=10^6$.\\
Bottom: The full line shows the chromatic number of large random 
graphs vs. their connectivity $c$. The symbols give results of 
{\it smallk} for $N=10^3$, each averaged over 100 
samples.}
\label{fig:complexity}
\end{figure}

Unlike in the case of finite-connectivity $p$-spin-glasses or,
equivalently, random XOR-SAT problems \cite{RiWeZe,CoDuMo,MeRiZe}, the
existence of a solution $t<1$ is not sufficient for a non-trivial
$\rho(\eta)$ to exist. The latter appears suddenly at the dynamical
transition $c_d$, which can be determined to high precision using the
population dynamical algorithm. This solution does not imply
uncolorability, but the set of solutions is separated into an
exponentially large number of clusters.  The logarithm of their
number, divided by the graph size $N$, is called the complexity
$\Sigma(c)$ and can be calculated from $\rho(\eta)$, cf. 
\cite{MeZe}
\begin{eqnarray}
\label{eq:Sigma}
\Sigma(c) &=& e^{-c} \sum_{d=1}^\infty \frac{c^d}{d!} \int
d\eta_1\cdots d\eta_d\ \rho(\eta_1)\cdots \rho(\eta_d) \nonumber\\
&&\times\ln \left(\sum_{l=0}^{q-1} (-1)^l {q \choose l+1} 
\prod_{i=1}^d [1-(l+1)\eta_i] \right) \nonumber\\
&& -\frac c2 \int d\eta_1 d\eta_2\ \rho(\eta_1) \rho(\eta_2)
\ln( 1- q \eta_1\eta_2 )\ .
\end{eqnarray}
The full derivation will be given in \cite{ColoringAlgo}.  At the
dynamical threshold, this complexity starts discontinuously with a
positive value, see Fig.~\ref{fig:complexity}, and decreases when $c$
is increased. The static RSB transition, and thus the $q$-COL/UNCOL
threshold $c_q$, are given by the point of vanishing complexity. At
this point the number of clusters becomes sub-exponential, and
disappears beyond $c_q$.

In the following table, we present the results for $q=3,4,$ and 5. For
the dynamical transition we show the corresponding values of $c_d$, of
the entropy $s(c_d)=\ln q + \frac c2 \ln \frac{q-1}q$ and the
complexity $\Sigma(c_d)$.  For the $q$-COL/UNCOL transition, the
critical connectivity $c_q$ and the solution entropy are given. Like
in random 3-satisfiability \cite{MoZe} and vertex covering
\cite{WeHa}, this entropy is found to be finite at the transition
point.

\begin{center}
\begin{tabular}{|c||c|c|c||c|c|}\hline
q & $c_d$ & $s(c_d)$ & $\Sigma(c_d)$ & $c_q$ & $s(c_q)$  \\ 
\hline \hline
3 & 4.42  & 0.203    & 0.0223  & 4.69 &  0.148  \\  
\hline
4 & 8.27  & 0.197    & 0.0553  & 8.90 &  0.106  \\  
\hline
5 & 12.67 & 0.196    & 0.0794  & 13.69 & 0.082  \\  
\hline 
\end{tabular}
\end{center}

One can see that the complexity at the dynamical threshold grows
strongly with $q$, whereas the total entropy
decreases slowly. This means that the clustering phenomenon becomes
more and more pronounced, the number of clusters increases, their
internal entropy $s(c)-\Sigma(c)$ becomes smaller.  It also becomes
more relevant for small systems. At $N=100$ and the dynamical
threshold, we would predict only around 10 clusters for $q=3$, for
$q=4$ this number would already be close to 250, and grow to about
2800 for 5 colors.

The dynamical transition is not only characterized by a sudden
clustering of ground states, at the same point an exponential number
of meta-stable states of positive energy appears \cite{MeZe}. Such
states are expected to act as traps for local search algorithms
causing an exponential slowing down of the search process. Well known
examples of search processes that are overwhelmed by the presence of
excited states are simulated annealing or greedy algorithms based on
local information.

To test this prediction, we have applied several of the best solvers
for COL and SAT problems available in the net
\cite{satlib,Culberson_HP}. The best results could be obtained using
the complete {\em smallk} program~\cite{Culberson_HP} which may need
exponential time to find a proper minimal coloring. Using a cutoff
time (we probed with 10 seconds, 1 minute and 2 minutes without
substantial changes for $N=10^3$), the algorithm can be restricted to
sub-exponential times, i.e. only the underlying polynomial-time
heuristic is applied. The results in Fig. 1 were obtained in the
following way: We first tried to color a random graph ($N=10^3$) with
a small number of colors (here $q=3$). If, after the cutoff time, {\it
  smallk} did not find a coloring, we stopped and retried with larger
$q$.  For each connectivity we averaged over 100 samples. As it can be
clearly seen, the algorithm fails with $q$ colors slightly below the
dynamical transition, confirming our expectations. Only a perfect
local heuristic should reach this threshold.

We conclude by noticing that, in similarity to the 3-SAT problem
\cite{MeZe}, we expect the assumptions underlying the cavity approach
to hold for single instances of COL. The equations for the order
parameter on single instances provide the full histogram of the N
probability distributions of the effective fields, one for each
variable, which describe the fluctuations of the polarization of each
Potts variable in the ground states.  On the physics side, this
information allows to develop a single sample statistical mechanics
analysis whereas on the algorithmic side it allows develop new
algorithms~\cite{ColoringAlgo}.

\begin{acknowledgments}
We are grateful to A. Braunstein, J. Culberson and F. Ricci-Tersenghi 
for interesting discussions.
\end{acknowledgments}


\begin{thebibliography}{99}
\bibitem{note:rsb} To test the correctness of the 1RSB solution we
  have numerically checked its local stability using the population
  dynamics described below. Further steps of RSB are expected to
  appear only beyond the COL/UNCOL transition due to a local
  instability.
\bibitem{note} This equation holds when the rewaiting parameter
$y$ is send to $\infty$ which is correct only inbetween the
dynamic and the static transition, cf. \cite{MeZe}. This is,
however, the case we are interested in.
\end{thebibliography}
\end{document}